\documentclass[12pt]{iopart}
\usepackage{graphicx}
\usepackage{cite}

\begin{document}
\title[Nonequilibrium wetting of finite samples]{Nonequilibrium wetting of finite samples}
\author{Thomas Kissinger, Andreas Kotowicz, Oliver Kurz,\\ Francesco Ginelli and Haye Hinrichsen}
\address{Fakult{\"a}t f\"ur Physik und Astronomie, Universit{\"a}t W{\"urzburg}
         \\D-97074 W{\"u}rzburg, Germany}

\begin{abstract}
As a canonical model for wetting far from thermal equilibrium we study a Kardar-Parisi-Zhang interface growing on top of a hard-core substrate. Depending on the average growth velocity the model exhibits a non-equilibrium wetting transition which is characterized by an additional surface critical exponent $\theta$. Simulating the single-step model in one spatial dimension we provide accurate numerical estimates for~$\theta$ and investigate the distribution of contact points between the substrate and the interface as a function of time. Moreover, we study the influence of finite-size effects, in particular the time needed until a finite substrate is completely covered by the wetting layer for the first time. 
\end{abstract}

\parskip 2mm


\def\xvec{\vec{x}}                      
\def\nupar{{\nu_\parallel}}             
\def\nuperp{{\nu_\perp}}                

\section{Introduction}
\label{Introduction}

When a chemically inert surface is exposed to a gas phase particles may
preferentially attach to the surface, forming a wetting layer of a
different phase. By changing physical parameters such as pressure or
temperature these systems may undergo a wetting transition at which the
thickness of the layer diverges and becomes macroscopic. In many experimental
situations it is reasonable to assume that the wetting layer is in thermal
equilibrium with the substrate so that methods of equilibrium statistical
physics can be applied~\cite{Dietrich86,EquilibriumFieldTheory}. More
recently, however, wetting phenomena far from equilibrium attracted considerable
attention~\cite{KPZWall,Wetting1,Wetting2,GiadaMarsili,Candia,MunozWetting,Wetting3}.
These studies are motivated by the question whether wetting processes under
non-equilibrium conditions, for example caused by a flux of particles or
energy, may show physical properties that are qualitatively different from
those of equilibrium wetting, especially at or in the vicinity of a wetting
transition. 

In order to describe a wetting process far from equilibrium one has to resort
to a {\it dynamical} description of the microscopic processes. This can be
done either by introducing explicit solid-on-solid growth
models~\cite{Wetting1,Wetting2} where the microscopical dynamics is described by
appropriate rules for deposition and absorption of adatoms over the substrate, 
or by studying appropriate Langevin equations
for non-equilibrium growth~\cite{MunozWetting}. Concerning their universal
properties all these studies were so far related to the Kardar-Parisi-Zhang
equation~\cite{KPZ} in one spatial dimension 
with an additional potential term~\cite{KPZWall} 
\begin{equation}
\label{LangevinEquation}
\frac{\partial h(\xvec,t)}{\partial t} = v_0 + \sigma\nabla^2 h(\xvec,t) -
\frac{\partial V\bigl(h(\xvec,t)\bigr)}{\partial h(\xvec,t)}
 \noindent + \lambda \bigl(\nabla h(\xvec,t)\bigr)^2 + \zeta(\xvec,t)\,,
\end{equation}
where $h(\xvec,t)$ is the height of the interface defining the thickness of
the growing layer, $v_0$, $\sigma$ and $\lambda$ are real parameters 
and $\zeta(\xvec,t)$ is a
white Gaussian noise. Here $V(h)$ is a potential that takes the influence of
the substrate into account. The nonequilibrium properties of this equation are
caused by the non-linear term $\lambda (\nabla h)^2$ which can be shown to be
a relevant perturbation in the renormalization group sense that violates
detailed balance. Obviously this term breaks reflection symmetry $h \to -h$ so
that in combination with a symmetry-breaking substrate the sign of $\lambda$
is expected to play a significant role. In the special case $\lambda=0$,
however, it can be shown that the dynamics in a stationary state obeys
detailed balance and is thus at thermal equilibrium. Hence by varying
$\lambda$ we can study the crossover from equilibrium to nonequilibrium
wetting. 

Without a substrate, i.e., for $V=0$, a one-dimensional interface evolving
according to the KPZ equation is known to roughen as $w(t) \sim t^\beta$,
where $w(t)$ is the width (defined as the standard deviation of the heights)
and $\beta$ is the so-called growth exponent. In a finite system of size $L$
this roughening process continues until the width saturates at $w_{\rm s}\sim
L^\alpha$ when the intrinsic correlation length $\xi \sim t^{1/z}$ reaches the
system size. Here $z$ and $\alpha=z\beta$ are the dynamical exponent and the
roughness exponent, respectively. In the case of KPZ growth in one spatial
dimension, where a fluctuation-dissipation theorem holds,
the exponents are given by simple rational values~\cite{Barabasi},
see Table~\ref{TABEX}. 

\begin{table}
\begin{center}
\begin{tabular}{|c||c|c|c|c|}
\hline             & $\alpha$ & $\beta$ & $z$    & $\theta$ \\ 
\hline $\lambda<0$ & \  $1/2$ \   &  \ $1/3$ \  & \ $3/2$ \  &  1.184(10)\\ 
       $\lambda=0$ &    $1/2$     &    $1/4$    &   $2  $    &  3/4\\ 
       $\lambda>0$ &    $1/2$     &    $1/3$    &   $3/2$    &  0.228(5)\\ 
\hline 
\end{tabular} 
\caption{
\label{TABEX} Critical exponents of wetting transitions in one spatial dimension}
\end{center}
\end{table}

In order to study nonequilibrium wetting, one has to introduce a substrate by
imposing a boundary in the microscopic dynamics or by choosing an appropriate 
form of the potential appearing in Eq. (\ref{LangevinEquation}). 
The simplest way to introduce such a boundary is to impose the restriction
$h(\xvec,t)\geq 0$, corresponding to an infinite-step potential of the form 
\begin{equation}
V(h)= \left\{
\begin{array}{ll}
0 & \mbox{ if } h\geq 0 \\
\infty & \mbox{ if } h<0
\end{array}
\right.\,.
\end{equation}
In order to avoid singularities in the continuous
formulation, this infinite-step potential is usually 
replaced in Eq. (\ref{LangevinEquation}) by an exponential function
\begin{equation}
\label{ExponentialV}
V(h) = \exp(-h)
\end{equation}
that was found to display essentially the same physical properties. 

Models defined by the addition of this simple boundary to a KPZ roughening
surface are known to exhibit a continuous wetting transition. In an infinite
system the critical point of this transition is determined by the asymptotic
average velocity $v_\infty=\lim_{t\to\infty}v(t)$ of a {\em free}
interface. For $v_\infty>0$ the interface moves away from the substrate so
that it no longer influences the dynamics. For $v_\infty<0$, however, the
interface is driven towards the substrate, approaching a steady state of
finite roughness that is characterized by a certain height profile. At the
transition point, where $v_\infty=0$, a scale-invariant behavior is
observed. Obviously, the order parameter that characterizes this transition is
the density of sites where the interface touches the substrate 
\begin{equation}
\rho_s(t) = \frac{1}{L} \sum_{i=1}^L \delta_{h_i(t),0}.
\label{orderp}
\end{equation}
or -- equivalently -- the spatial average $\overline{\exp(-h(\xvec,t))}$ in
the continuous formulation.  

\begin{figure}
\centerline{\includegraphics[width=160mm]{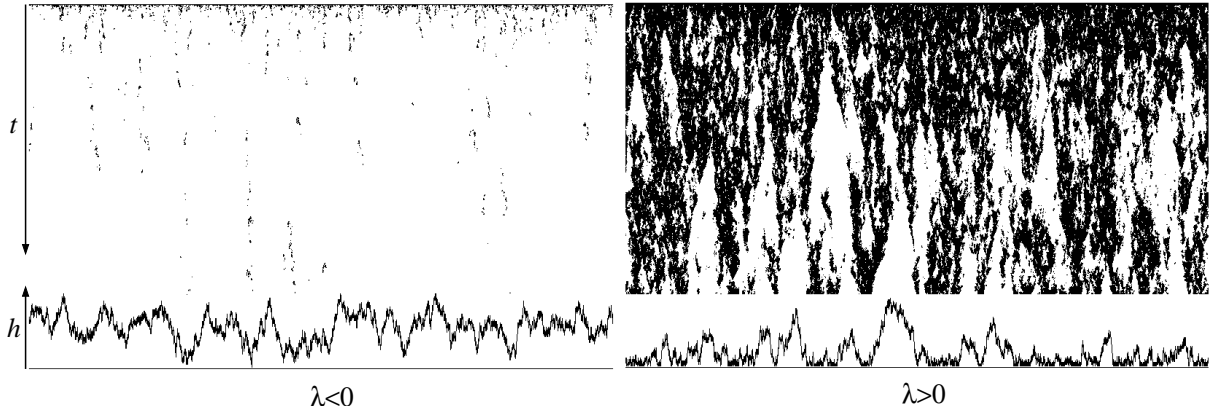}}
\caption{\label{FIGDEMO} \small
Upper row: Typical spatio-temporal pattern of contact points (pinned sites) between interface
and substrate in the SSM (see text) at criticality, starting with a flat interface at
$t=0$. Lower row: Final configuration of the interface at $t=5000$.  
Note that for $\lambda<0$ there are only few contact points, while for
$\lambda>0$ the density of contact points is much higher. 
}
\end{figure}

From a theoretical point of view the influence of a hard-core substrate
without additional short-range forces can be interpreted as a boundary condition in the
space spanned by $\xvec$, $t$, and $h$. As in other critical phenomena, this
boundary condition does not change the underlying universality class and the
associated bulk exponents, meaning that the process still belongs to the KPZ
universality class. However, the substrate as an additional boundary condition
adds a new feature, characterized by an order parameter $\rho_s(t)$ which is
associated with additional surface exponent $\theta$. This order parameter was found
decrease with increasing time as a power law~\cite{KPZWall} 
\begin{equation}
\label{SurfaceScaling}
\rho_s(t) \sim t^{-\theta}\,.
\end{equation}
The exponent $\theta$ is expected to take three different values depending on the sign of
the parameter $\lambda$ of the nonlinear term. 
In fact, already a visual inspection of typical interfaces at
criticality reveals significant differences depending on the sign of
$\lambda$. As can be seen in Fig.~\ref{FIGDEMO}, for $\lambda<0$ only few
contact points remain, leading to a fast decay of $\rho_s(t)$, while for
$\lambda>0$ the density of contact points decays only slowly.  

Quite interestingly, it has recently been discovered, both through numerical
simulations and analytical arguments \cite{Baroni, Pikovsky}, that also the 
{\it complete synchronization} phase transition (CST) in most spatially extended
chaotic dynamical systems belongs to the same universality class of a bounded
KPZ equation with a negative nonlinear term $\lambda$. 
It can be shown that, when the local dynamic is smooth enough 
(more precisely: the linearization of the system around the synchronized
state correctly describes the propagation of {\it finite} perturbations),
the critical behavior of the CST is described by the so called 
{\em multiplicative noise} (MN) Langevin equation
\begin{equation}
\label{MN}
\frac{\partial}{\partial t} n(\xvec,t) =
a n(\xvec,t)-n^2(\xvec,t)+D \nabla^2 n(\xvec,t) + n(\xvec,t) \zeta(\xvec,t)
\end{equation}
where the order parameter $n(\xvec,t) \geq 0$ is the coarse grained density 
of unsynchronized
regions. Here $a$ and $D>0$ are real parameters, and 
$\zeta(\xvec,t)$ is the same noise as in Eq.~(\ref{LangevinEquation}).
By a Cole-Hopf transformation 
\cite{KPZWall} 
\begin{equation}
h(\xvec,t) -\log n(\xvec,t) 
\label{HopfCole}
\end{equation}
Eq. (\ref{MN}) turns exactly into the bounded KPZ equation (\ref{LangevinEquation})
with $\lambda = -\sigma = -D$ and an exponential potential as in Eq.
(\ref{ExponentialV}). This indicates that the MN equation
is just a realization of the KPZ universality class and its
non-trivial surface critical behavior.
In particular the order parameter $n(\xvec,t)$ in Eq.~(\ref{MN}) 
scales exactly in the same way as the density
of contact points $\rho_s(t)$ and therefore its spatial average will decay as
$t^{-\theta}$. 

Eq. (\ref{MN}) has been studied numerically in a
series of papers~\cite{MN1,MN2, Chate} and its transition is said to belong to
the multiplicative noise I (MN1) class.
Moreover, the related multiplicative Langevin equation which 
can be derived via an inverse Cole-Hopf transformation from the bounded KPZ equation 
(\ref{LangevinEquation}) with $\lambda>0$ has been numerically analyzed
in Ref. \cite{MunozRev}. The associated transition is said to belong to the 
multiplicative noise II (MN2) class.

Presently, no exact analytical derivation of the surface exponent $\theta$ in
the nonequilibrium case is known. While in Ref. \cite{MF} the surface exponent
has been obtained for both the MN1 and MN2 classes 
with a mean-field like approximation suited for one spatial dimension, 
the presence of a strong-coupling fixed point and of essential divergences
makes the exact values of the surface exponent
not accessible by known renormalization group techniques.
On the other hand, as we shall see, 
the equilibrium case $\lambda=0$, sometimes referred
to as bounded Edward-Wilkinson, yields the simple rational result for 
the surface exponent $\theta_{EW} = 3/4$. 
In light of this result, it would be desirable to know if also the 
nonequilibrium surface exponents 
have simple rational values, an indication that the surface critical behavior
is a trivial consequence of the known KPZ bulk properties. In particular,
in Ref. \cite{DrozPaper} the value $\theta = 7/6$ has been 
conjectured for $\lambda<0$. Moreover, it is not 
yet fully clear how the contact points between
interface and substrate are distributed and how finite-size effects influence
the dynamics at the substrate. 

As a first step toward a better comprehension of the bounded KPZ class, we present
here a detailed analysis of a simple solid on solid model generally belonging
to the KPZ universality class,
the so-called single step model (SSM). Once provided with an additional
boundary restriction without additional short-range forces, 
it shows a simple wetting transition belonging to the bounded KPZ
universality class \cite{Ginelli03}. In particular, the simplicity of the
model allows for an analytical knowledge of it's wetting transition critical point
$v_{\infty}=0$.  In Sect.~\ref{SSM} the efficiency of its algorithm 
allows us to improve current numerical estimates of the exponent~$\theta$. 

In Sect.~\ref{GAP} we investigate the distribution of intervals between the
contact points, identifying significantly different scaling behaviors
depending on the sign of $\lambda$. Finally, in Sect.~\ref{GAP2}, 
we study the scaling properties
of \textit{finite} systems, analyzing the time that is needed before the
interface detaches globally for the first time. This kind of finite size
scaling analysis is a common tool in numerical studies of nonequilibrium
systems with an absorbing state, such as Directed Percolation (DP) \cite{DP}, where
the average absorbing time is known to scale according to the dynamical
exponent, $\tau \sim L^z$. However, supported by numerical results
and analytical arguments, we argue that in the case $\lambda \leq 0$
this detachment time does {\em not} grow as a power law with the system size. 
Conclusions are drawn in Sect. \ref{CONC}, while
in Appendix A we analytically compute the surface exponent when detailed
balance is imposed in the SSM, thus proving the equilibrium result
$\theta_{EW} = 3/4$.

\section{Measurement of the surface exponent in the single step model}
\label{SSM}

\begin{figure}
\centerline{\includegraphics[width=125mm]{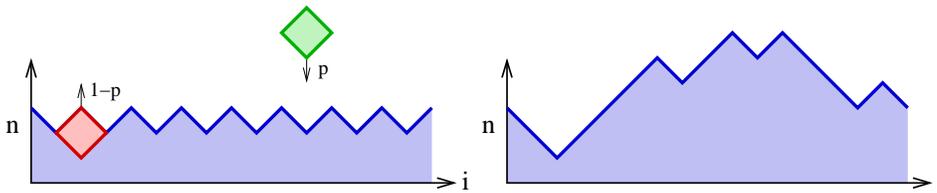}}
\caption{\label{FIGSSM} \small
Single step model in 1+1 dimensions. Left panel: The simulation starts with
a flat interface in the form of a horizontal sawtooth pattern.
On selecting a site at a local minimum
a diamond (rhombus) is deposited with probability~$p$, flipping up the interface
by two units. Similarly, selecting a site at a local maximum, a diamond 
evaporates with probability $1-p$, flipping the interface downward by two units.
Right panel: Roughening interface after several deposition and evaporation events.
}
\end{figure}

In what follows let us consider the so-called single step model (SSM) in one
spatial dimension, which is probably the simplest and most compelling lattice
model for KPZ-type interface growth. In this model the growing interface is
represented by a set of integer heights $h_i$ residing at the sites $i=1\ldots
L$ of a one-dimensional lattice of length $L$ with periodic boundary
conditions, obeying the restriction 
\begin{equation}
\label{restriction}
h_{i+1}-h_i=\pm 1\,.
\end{equation}
The model is controlled by a single parameter $p \in [0,1]$ and evolves by
random sequential updates as follows: Choosing a random site $i$ the height
$h_i$ is increased by two units with probability $p$ provided that the
restriction (\ref{restriction}) is not violated. As shown in Fig.~\ref{FIGSSM}
this move can be thought of as depositing a diamond. Similarly, the height
$h_i$ is decreased by two units with probability $1-p$ provided that the
restriction (\ref{restriction}) is not violated. As usual each local update
attempt corresponds to a time increment of $dt=1/L$.

For $p=1/2$ the propagation velocity of the interface is zero, detailed
balance holds in the stationary state and its dynamics is
effectively described by an Edwards-Wilkinson equation~\cite{EW}. For $p \neq
1/2$, however, the SSM exhibits KPZ growth with $\lambda$ being proportional
to $\frac12-p$. In this case the propagation velocity is nonzero and depends
on the roughness and the average slope of the interface. For example,
the initial velocity of a flat interface (a sawtooth pattern as shown in the
left panel of Fig. \ref{FIGSSM}))
is given by 
\begin{equation}
v(0) = 2p-1\,.
\end{equation}
As time proceeds, the interface roughens and the propagation velocity
decreases until it saturates at a certain value
$v_L(\infty)$ which depends on the system size. For the KPZ class the
so-called excess velocity $v_L(t)-v_L(\infty)$ is known to decay with time as
$t^{-2/3}$~\cite{Krug}. As a major breakthrough, Pr{\"a}hofer and Spohn recently succeeded in 
computing the rescaled height profile of a roughening KPZ 
analytically~\cite{PraehoferSpohn}.

In contrast to many other KPZ growth models the SSM allows one to calculate
the asymptotic growth velocity $v_L(\infty)$ exactly. To this end one
identifies segments of positive (negative) slope with particles (vacancies),
mapping the single step model onto a partially asymmetric exclusion process
(ASEP)~\cite{ASEP} of diffusing particles with density $1/2$. Since the ASEP
is known to evolve towards an uncorrelated product state with a current
$j=p/2-1/4$, one can easily show that the propagation velocity of the
interface in a finite system of size $L >> 1$ tends to 
\begin{equation}
v_L(\infty) = \lim_{t \to \infty} v_L(t) = \left(p-\frac12 \right) \left(1+\frac{1}{L}\right)\,.
\end{equation}
In order to study nonequilibrium wetting we need to introduce a hard-core
substrate. As there is no parameter in the SSM to control the asymptotic growth velocity
of the interface independently from the KPZ nonlinear term $\lambda$,
the substrate itself has to move so that we can approach the
wetting transition by tuning its velocity $v_s$. Obviously, this transition
takes place even in finite systems and the critical point is exactly given by
$v_s=v_L(\infty)$.  

To determine the surface exponent $\theta$ we simulated the SSM for $p=1/2$ as
well as for $p=0,1$, where the non-linear effects of the KPZ term are most
pronounced. In order to minimize finite-size effects we use a very large
lattice size $L=10^6$. For $p=1/2$ (corresponding to the equilibrium case
$\lambda=0$ in the KPZ equation) the interface velocity at the transition
vanishes so that the substrate can be implemented by rejecting all updates
that would lead to a negative height. The process is then simulated as usual
and the density of sites at zero height is averaged over many independent
runs. For $p=1$ (corresponding to $\lambda<0$) we advance the substrate in
temporal intervals $\Delta t=2(1-1/L)$ by one unit, flipping all sites below
the new base line upward. Finally, for $p=0$ (corresponding to $\lambda>0$)
the substrate moves backward by one unit every interval $\Delta t$, rejecting
all updates that would render a height below the actual base line. Since the
moving baseline effectively introduces a discrete time step in the 
algorithm, the density of contact points is measured 
\textit{immediately after (before)} moving the wall when $p=1$ ($p=0$).

%
%
\begin{figure}
\includegraphics[width=155mm]{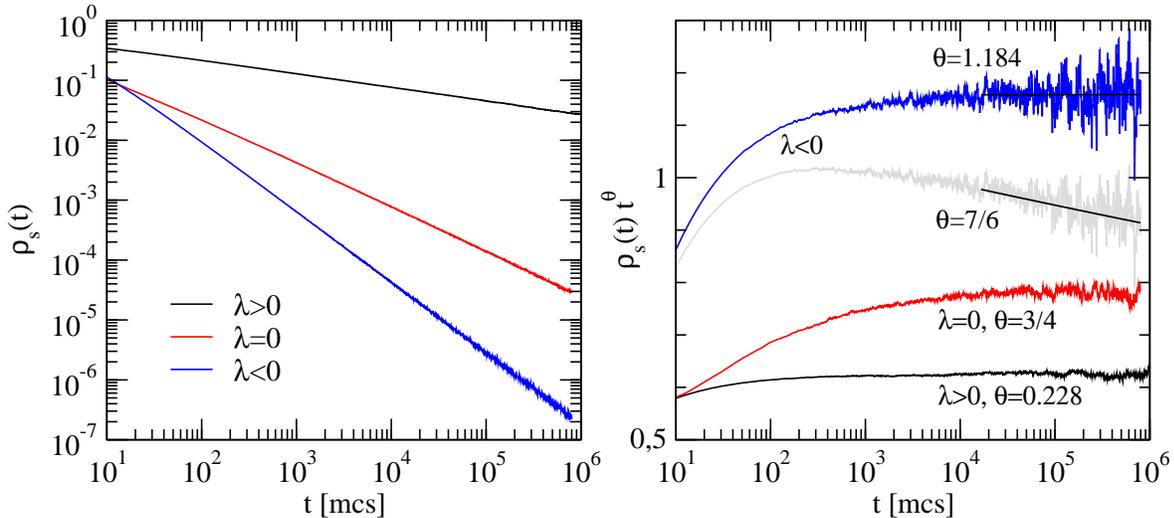}
\caption{\footnotesize
\label{FIGTHETA}
Measurement of the exponent $\theta$ in the single step model. The left panel
shows the density of contact points $\rho_s(t)$ as a function of time for
$p{=}1$ $(\lambda{<}0)$, $p{=}1/2$ $(\lambda{=}0)$, and $p{=}0$
$(\lambda{>}0)$. The right panel shows the same data multiplied with the
estimated power law $t^\theta$. In addition we plotted the curve for $p{=}1$
$(\lambda{<}1)$ multiplied by $t^{7/6}$, demonstrating that the exponent
$\theta=7/6$ conjectured by Droz and Lipowski is incompatible with the
numerical data. 
}
\end{figure}
%
%
The numerical results are shown in Fig.~\ref{FIGTHETA}. Simulating up to
$8\times 10^5$ time steps and averaging over $160$ independent runs one
obtains convincingly straight lines in a double-logarithmic plot. Extrapolating
local slopes we arrive at the estimates 
\begin{equation}
\theta = \left\{
\begin{array}{ll}
0.228(5) & \mbox{ for } p=0    \\
0.750(5) & \mbox{ for } p=1/2  \\
1.184(10) & \mbox{ for } p=1 
\end{array}
\right.
\end{equation}
In order to illustrate the influence of scaling corrections, we multiplied
each data set with~$t^\theta$ in the right panel of Fig.~\ref{FIGTHETA} so
that the curves become horizontal for large $t$. 
Our results have to be compared to the early numerical estimates 
$\theta_1 = 1.1(1)$ and $\theta_2 = 0.215(15)$ for the MN1 and MN2 
Langevin equations~\cite{MunozRev}. A more recent numerical study of the MN1 equation
yielded $\theta_1 = 1.21(3)$ \cite{Chate}.

The numerical result for $p=1$ is particularly interesting as it seems to rule
out an earlier conjecture by Droz and Lipowski~\cite{DrozPaper}. Analyzing a
synchronization transition between two coupled lattices of tent maps and
making use of a hyperscaling relation between stationary and spreading
exponents, they proposed that for $\lambda<0$ the
exponent $\theta$ should be given by the exact value $7/6 \approx
0.1666$. With the accuracy of the present simulations, however, this value
lies clearly outside the error margin. To illustrate this difference we plotted
$\rho_s(t)\,t^{7/6}$ in the right panel of Fig.~\ref{FIGTHETA}. As can be
seen, the curve has a negative slope over at least two decades before
finite-size effects set in at $t \approx 10^5$. Our result suggests that in
both cases of $\lambda \neq 0$ the exponent $\theta$ is probably \textit{not} 
a simple rational number, rather it seems to be irrational. 
For $\lambda=0$ ($p=1/2$),
however, our numerical estimate is in fair agreement with $\theta=3/4$. In
fact, as shown in Appendix A this value can be proven
analytically provided that the two-point $C(\ell)$ function of contact points
decays algebraically as $\ell^{-\theta z}$. 

\section{Gap distribution of an infinite system}
\label{GAP}

Let us now investigate how the contact points of a critical interface with the
substrate are distributed in an infinite system at a given time. Here an
important quantity is the {\em gap distribution} $G(\ell,t)$ of distances
$\ell$ between neighboring contact points measured at time $t$. As
Fig.~(\ref{FIGDEMO}) suggests, this distribution depends significantly on the
sign of $\lambda$.  

%
%
\begin{figure}
\includegraphics[width=160mm]{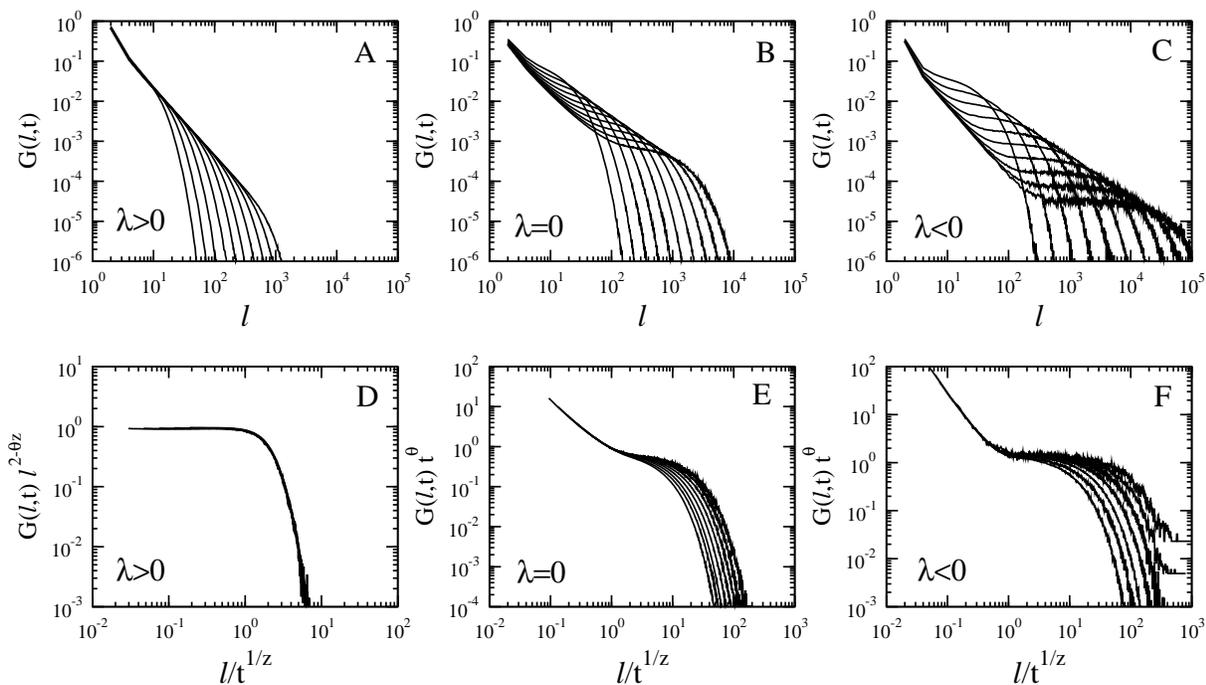}
\caption{\footnotesize
\label{FIGGAPS}
Panels A-C show the gap distribution function depending on the gap size
measured in the SSM for $p=0,1/2$, and 1 at fixed times $t=16,32,64,\ldots
8192$. Panel~D shows a collapse of the data for $\lambda<0$ according to the
scaling form~(\ref{ScalingForm}). Contrarily, panels E and F demonstrate that
this collapse does not work for $\lambda \leq 0$. 
}
\end{figure}
%
%
Our numerical results are shown in Fig.~\ref{FIGGAPS}. For $p=0$ ($\lambda>0$)
one observes an ordinary algebraic decay with a cutoff at the correlation
length $\xi \sim t^{1/z}$. This type of decay is the same is in other
non-equilibrium critical phenomena as, for example, in DP. For
$p\geq 1/2$ ($\lambda\leq 0$), however, the curves first decay algebraically,
then reach a {\it plateau}, and finally cross over to an exponential
cutoff. 

The qualitative difference between these two cases is due to the
presence of two different length scales, namely, the bulk correlation length
$\xi$ and the average gap size $\overline{\ell}$. These two length scales grow
with time as 
\begin{eqnarray}
\xi(t) &\sim &  t^{1/z} \\
\overline{\ell}(t) &\sim & t^{\theta}\,,
\end{eqnarray}
the latter equation being a consequence of the relation $\overline{\ell}(t) =
\frac{1}{\rho_s(t)}$ which holds for any point distribution on a
one-dimensional line\cite{Ginelli05}. 
Depending on the values of the exponents $z$ and $\theta$, two different
scenarios are encountered: 
\begin{itemize}
\item
For $\lambda>0$ $(p<1/2)$ we have $\theta<1/z$ and thus the average distance
between contact points $\overline{\ell}(t)$ is \textit{smaller} than the
correlation length $\xi(t)$. In this case the gap distribution is a simple
power law with an exponential cutoff at $\xi(t)$. As in other critical
phenomena this suggests the usual scaling form 
\begin{equation}
\label{ScalingForm}
G(\ell,t) \sim \ell^{-(2-\theta z)} \, f\left( \frac{\ell}{t^{1/z}} \right).
\end{equation}
In fact, plotting $G(\ell,t) \ell^{2-\theta z}$ versus $\ell/t^{1/z}$ one
obtains a convincing data collapse, as shown in Fig.~\ref{FIGGAPS}D.\\ 

\item
For $\lambda\leq 0$ $(p \geq 1/2)$, where $\theta>1/z$, the average distance
between contact points is \textit{larger} than the correlation length. In this
case the gap distribution function is found to decay initially as $\ell^{-\theta
 z}$. Reaching the bulk correlation length $\ell \approx \xi(t)$ it crosses
over to an exponential function. However, in contrast to the previous case the
associated length scale of this exponential function is $\overline{\ell}\sim
t^{\theta}$, hence in a double-logarithmic graph the curves exhibit a plateau
at level $\propto t^{-\theta}$ extending from $\xi(t)$ to
$\overline{\ell}(t)$, followed by an exponential cutoff. 
\end{itemize} 
In the following Section we will use these postulated scaling forms for the gap distribution
to derive the first detachment time in a finite system, showing that the two cases lead to 
very different types of finite-size scaling. We note that the power-law scaling of the
gap distribution is a postulate, supported by numerical results shown in Fig.~\ref{FIGGAPS}
and the generally accepted power-law scaling of $\rho_s(t)=1/\bar{\ell}(t)\sim t^{-\theta}$.
However, to our knowledge a rigorous proof of power-law scaling in the gap distribution 
is still missing.

\section{First detachment in a finite system}
\label{GAP2}

We finally turn our attention to finite size scaling analysis, which a common
tool to analyze properties of systems at criticality. 
We recall that the correct order parameter describing the wetting transition is
the density $\rho_s(t)$ of sites pinned to the substrate, as defined in
Eq. (\ref{orderp}). Comparison with DP naively leads one to draw an analogy
between the pinned sites and the infected sites of systems with an absorbing
state (i.e. a state in which the system dynamics get trapped forever). 
However, in the present case the depinned 
phase (i.e. the phase in which the substrate is completely
covered) is {\it not} absorbing in the DP sense. In fact, since the wetting
transition takes place at zero relative velocity, fluctuations can
easily bring the interface back to the substrate, thus leaving the depinned
state. Hence, instead of studying the average {\it survival} time $\tau_{abs}$ 
(i.e., the average time at which the system becomes trapped in an absorbing state), 
here we are forced to consider merely the {\it first} 
depinning time $\tau_{f}$, i.e., the average time at which an initially
exposed substrate gets completely covered for the first time.

While in systems with an absorbing phase one expects the average survival time
in a system of size $L$ to scale as $\tau_{abs} \sim L^z$, where $z$ is the
dynamical exponent, our numerical results will show an apparent power law
behavior $\tau_{f} \sim L^{\gamma}$, where the exponent $\gamma$ depends on the sign
of $\lambda$ and is generally different from the KPZ dynamical exponent.
However, a more detailed analysis of the first depinning time probability
distribution $P_L(t)$ will show that at least for $\lambda \leq 0$ 
the scaling form 
\begin{equation}
P_L(t) \sim L^{-\gamma} g(t/L^z)
\label{scaling2}
\end{equation}
is not satisfied, thus indicating that the observed power-law behavior does
not hold asymptotically. 
 
\subsection{Numerical results}
%
\begin{figure}
\includegraphics[width=152mm]{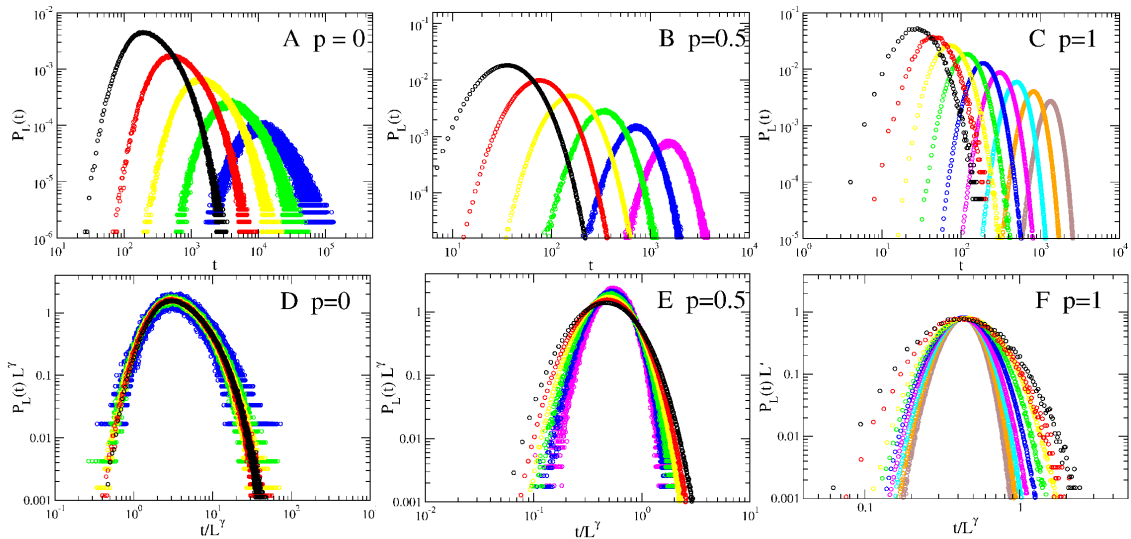}
\caption{\footnotesize
\label{FIGDPDISTRIBUTIONS}
Panels A-C show the distributions of the first depinning times for for p=0, 0.5 and 1.
The system size increases from left to right starting with L=64 and going up to L=1024, 2048 and 16384 respectively.
In panels D-F we test a collapse of the the above distributions according to
Eq. (\ref{scaling2}). Only for $p=0$ the collapse works convincingly.}\end{figure} 
In order to measure the first depinning time $\tau_f$, we used the single
step model with a moving hard-core substrate as described above. To test
finite size scaling we simulated various system sizes ranging from $L=64$ to
$L=16384$.

\begin{figure}
\includegraphics[width=150mm]{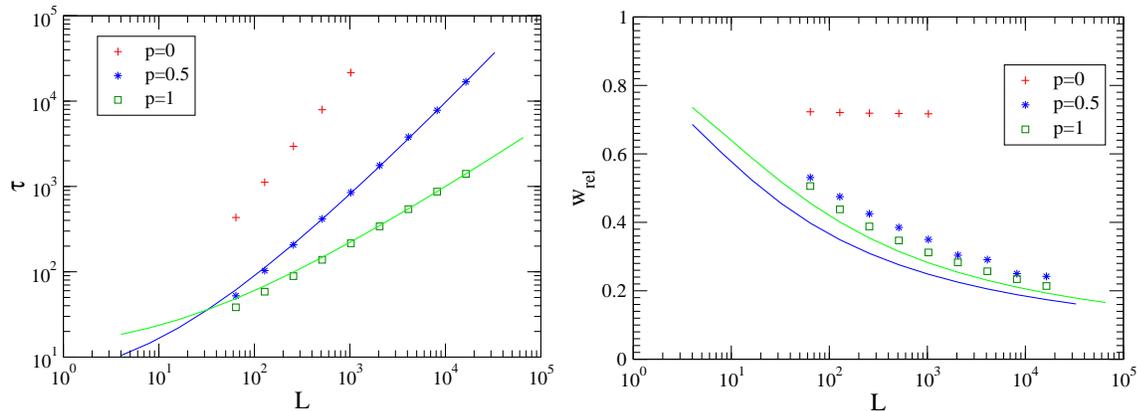}
\caption{\footnotesize
\label{FIGAVERAGEDP}
Left: Mean depinning time $\tau_f$ plotted against the system size L. Right: 
Relative width $\sigma_{\rm rel}$ of the distributions as a function of $L$. 
As can be seen, the relative width stays constant for $p=0$, in
contrast to the cases $p=0.5$ and $p=1$. The solid lines represent the
theoretical predictions for the cases $p=0.5$ (blue) and $p=1$ (green) 
according to Eq.~(\ref{TheoreticalResult}).} 
\end{figure}

For each realization we determined the time $t$ at which the interface was completely
depinned for the first time. The resulting histograms of the 
probability distribution $P_L(t)$ of
first depinning times are displayed in Fig.~\ref{FIGDPDISTRIBUTIONS}. Moreover, we
calculated the average depinning time for different system sizes and the
relative width of the characteristic distributions
\begin{equation}
\sigma_{rel} = \frac{\Delta \tau_{f}}{\tau_{f}}
\label{relw}
\end{equation}
where $\Delta \tau_{f}$ is the standard deviation of first depinning times.
The dependence of both quantities on $L$ is shown in Fig.~\ref{FIGAVERAGEDP}. 
Our numerical results can be summarized as follows:

\begin{itemize}
\item
In the case of $\lambda > 0$ we simulated the SSM with $p=0$ (i.e., no
absorption just desorption), averaging over $1.56 \times 10^6$ realizations 
for each system sizes up to $L=1024$. We find that there is a strong 
indication of an ordinary power law behavior. The average
depinning time scales with $L^{\gamma}$, with $\gamma=1.41(2)$, while
the probability distributions $P_L(t)$ have a constant
relative width and,
as it is shown in Fig.~\ref{FIGAVERAGEDP} (panel D),
they can be cleanly collapsed according to Eq. (\ref{scaling2}).\\
  
\item
To study the case $\lambda=0$ and $\lambda<0$ we simulated the SSM with 
$p=0.5$ and $p=1$ for system sizes ranging from $L=64$ to $L=2048$ ($p=0.5$)
or to $L=16384$ ($p=1$),
averaging over $8 \times 10^5$ realizations. In these cases too, the average
depinning time $\tau_{f}$ shows an apparent power law scaling. However, the
relative width $\sigma_{rel}$ of the probability distributions \textit{shrinks} 
with increasing system size so that the
probability distribution curves cannot be convincingly collapsed, 
making it obvious that we have no ordinary scaling behavior 
(see Figs.~\ref{FIGDPDISTRIBUTIONS} and~\ref{FIGAVERAGEDP}).
As it will be shown in the next section, this rather unconventional behavior 
can be traced back to the particularities of the gap distribution.
\end{itemize} 
%

\subsection{Analytical arguments for the case $\lambda \leq 0$}
\label{AnalyticalArguments}

\begin{figure}
\begin{center}
\includegraphics[width=125mm]{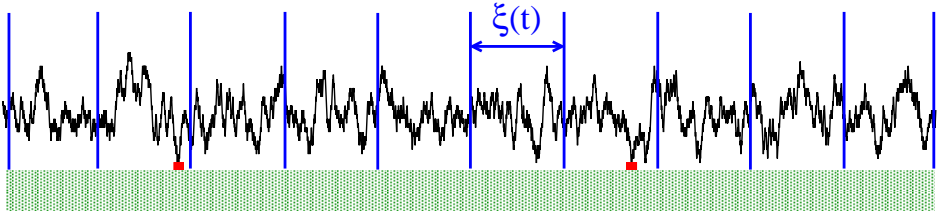}
\end{center}\caption{\footnotesize
\label{FIGDETACH}
Schematic sketch of a critical interface in a finite system for $\lambda\leq 0$ at a fixed time $t$, divided into  mutually uncorrelated segments of size $\xi(t)$. The contact points are marked as red dots. The moment of first detachment corresponds to a fluctuation where all segments independently detach from the surface for the first time.
}
\end{figure}

In the following we show how the unconventional behavior observed for
$\lambda \leq 0$ can be explained analytically  in terms of the gap
distribution function. The calculation is based on the observation that for
$\lambda \leq 0$ the moment of first detachment in a finite system
typically takes place long before the correlation length $\xi(t)$ reaches the
system size. This means that the interface may be thought of as being divided
into $L/\xi(t)$ \textit{uncorrelated} segments, each of them touching the
substrate with a small probability~(cf. Fig.~\ref{FIGDETACH}). The moment of
first detachment is then interpreted as the first fluctuation in which 
all these segments happen to detach independently. 
This assumption justifies the following approximations:

\begin{itemize}
\item[(a)]
The distribution of first detachment times in a finite sample of size $L$ with periodic boundary conditions approximately coincides with the corresponding distribution of first detachment for an arbitrarily chosen segment of size $L$ in an \textit{infinite} system.\\
\item[(b)]
The distributions of contact points at different times are assumed to be uncorrelated.
\end{itemize}
These two assumptions allow us to use the gap distribution investigated in Sect.~\ref{GAP} and to compute the distribution of first detachment times analytically. While assumption (a) is reasonable in systems much larger than the correlation length, assumption (b) is crucial as it neglects all temporal correlations of contact points. 

Obviously, the probability for a segment of size $L$ in an infinite system to detach at time $t$ is given by
\begin{equation}
Q^{d}_L(t) = \frac{\int_L^\infty d\ell\, (\ell-L) \, G(\ell,t)} {\int_0^\infty d\ell \, \ell \, G(\ell,t)}\,,
\end{equation}
where $G(\ell,t)$ is the gap distribution discussed in Sect.~\ref{GAP}. Consequently, the probability $Q^{n}_L(t)$ that the segment \textit{never} detached up to time $t$ is given by the product of $1-Q^{d}_L(t) \, dt$ over all time steps or -- in a continuous formulation -- by an exponentiated integral
\begin{equation}
\label{ExponentialApproximation}
Q^{n}_L(t) = \exp \left( -k \, \int_0^t \, dt' \, Q^{d}_L(t') \right) \,
\end{equation}
where $k$ is a constant to be fitted to the numerical data. The quantity of interest, namely, the probability of first detachment, is then given by
\begin{equation}
P_L(t) = -\,\frac{d}{dt}  Q^{n}_L(t) \quad.
\end{equation}
%
%
%
%
%
%
An examination of these expression reveals that the probability of first
detachment is determined by the occasional occurrence of very large gaps. This
means that $P_L(t)$ is governed by the right edge of the plateau
shown in Figs.~\ref{FIGGAPS}B-\ref{FIGGAPS}C while the initial power law decay
of the gap distribution has virtually no influence. Therefore, 
we may approximate the gap distribution function by
\begin{equation}
G(\ell,t) \simeq A\, t^{-\theta} \exp\left( -B\frac{\ell}{t^\theta}\right)
\end{equation}
where $A$ and $B$ are non-universal constants. 
With this ``exponential'' approximation the above integrals can be solved, giving
\begin{equation}
Q^{d}_L(t) =  \exp\left( -B\frac{L}{t^\theta}\right)
\label{expapprox}
\end{equation}
and
\begin{equation}
\label{TheoreticalResult}
P_L(t) = k \, \exp \left(
-B \frac{L}{t^\theta} - \frac{k (BL)^{1/\theta}}{\theta} \Gamma(-1/\theta,BL/t^\theta)
\right)\,,
\label{Pfirst}
\end{equation}
where $\Gamma(a,z) = \int_z^{\infty} t^{a-1} e^t \,dt$ is the incomplete gamma function.
Here the scaling-invariant combination $L/t^{\theta}$ appears in two
places whereas the prefactor of the gamma function is proportional to
$L^{1/\theta}$. Obviously it is this prefactor which breaks scaling invariance
so that $P_L(t)$ does not satisfy the scaling form (\ref{scaling2}).

%
%
\begin{figure}
\includegraphics[width=150mm]{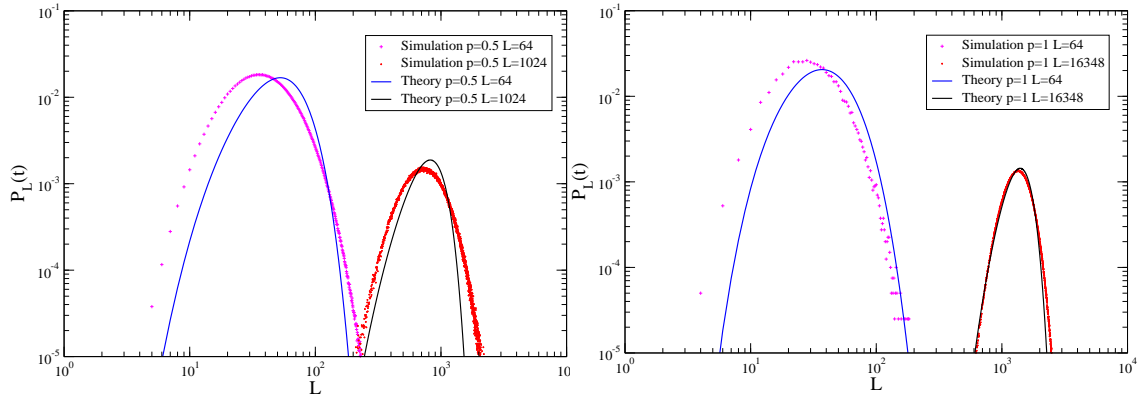}
\caption{\footnotesize
\label{FIGANALYTICAL}
Overlay of analytical predictions for the first depinning time probability distribution for $\lambda \leq 0$ ($p$=0.5 and 1) and numerical results of our simulation on selected system sizes.}
\end{figure}
%
%
In Fig.~\ref{FIGAVERAGEDP} the average depinning time and the relative width
computed numerically from Eq. (\ref{Pfirst}) for $p=1, 1/2$ are found to 
be in good agreement with numerical data, while 
Fig.~\ref{FIGANALYTICAL} shows a direct comparison between the analytical and
the numerically derived probability distribution functions for the first
depinning times. The parameters $B$ and $k$ have been determined by
fitting to the numerical data and to grant normalization.
Discrepancies between theory and simulations mainly occur
for small system sizes and can
be traced back to the above approximation of the gap distribution by a
exponential function in Eq.~(\ref{ExponentialApproximation}). 
As can be seen in Figs.~\ref{FIGGAPS}B-C the initial power-law part 
of the gap distribution function neglected in the theory is expected 
to play a more pronounced role for small system sizes. 
However, for large system sizes, where the ``exponential'' approximation
used to derive Eq. (\ref{expapprox}) performs better, 
the agreement steadily improves. 

\section{Conclusions}
\label{CONC}

Nonequilibrium wetting can be modeled by a KPZ growth process combined with a
hard-core substrate that gives rise to an additional potential term in the
KPZ equation. Adding such a term breaks up-down symmetry and
leads to  non-trivial surface features, which depend on the sign of the
parameter $\lambda$ of the KPZ nonlinear term. For numerical studies of 
nonequilibrium wetting transitions we advocate the use of the single-step model 
because of its simplicity and efficency. 

With large-scale simulations presented in this paper we have improved 
numerical estimates of the surface exponent $\theta$. These estimates 
strongly suggest that $\theta$ takes non-trivial irrational values for
$\lambda \neq 0$. In particular we can rule out the simple rational 
value $\theta=7/6$ for $\lambda <0$ suggested by Droz and Lipowski. The
failure of their argument can probably be traced back to the
absence of an absorbing state in the bounded KPZ transition, questioning
the feasibility of spreading analysis and hyperscaling relations used
in absorbing phase transitions in the present case. However, 
in the equilibrium case ($\lambda=0$) our result confirm the simple
rational value $\theta = 3/4$, in agreement with analytical arguments.

In order to investigate finite-size scaling of wetting processes described by
the bounded KPZ equation we introduced the average first depinning time 
as opposed to the average survival time in absorbing phase transitions. 
In one spatial dimension this naturally leads 
one to consider the distribution of gaps between pinned sites. 
We find that two different length scales are shaping the surface dynamics, namely,
the spatial correlation length $\xi(t) \sim t^{1/z}$ and the average gap length 
$\bar{\ell}(t)\sim t^\theta$. If this latter length is larger then $\xi(t)$, 
as in the case $\lambda \leq 0$, gaps between contact points effectively behave as
uncorrelated, thus breaking scale invariance for the first depinning times.
This result comes as a surprise if one naively considers the analogy with
absorbing phase transitions, where the scaling between temporal and spatial
quantities is simply dictated by the dynamical exponent $z$.

The case of $\lambda > 0$ on the other hand, shows a clean finite size scaling behavior
of the first depinning times. Since in this latter case the spatial
correlation length is larger than the average gap size, the analytical
arguments presented in Sect.~\ref{AnalyticalArguments} no longer apply so that
we are not able to predict the shape of the first depinning time 
probability distribution $P_L(t)$. However, we note that our numerical result
$\gamma=1.41(2)$ is less then $10\%$ off the known value for the KPZ dynamical
exponent $z=1.5$. Since it has been proven in~\cite{MunozRev} that spatial and 
temporal correlations are not asymptotically altered by the additional boundary
condition, we expect $\gamma$ to converge gradually versus $z$ in the limit
$L\to\infty$. However, for large but finite systems within the numerically 
accessible range the estimates may deviate. 
 
%
\noindent
{\bf Acknowledgments:}\\
We are grateful to A. Torcini, A. Pikovsky and C. Godr{\`e}che for fruitful discussions.
Most of the numerical simulations were carried out on the 80-node CIP cluster 
at the University of W{\"u}rzburg. We would like to thank A. Klein and A. Vetter for excellent technical support.

\appendix
\section{Derivation of the exponent $\theta$ in the equilibrium case}
\label{APPTHETA}
In this Appendix we derive the exponent $\theta$ for the equilibrium case $p=1/2$ in the SSM analytically. In this case all transition rates between mutually reachable interface configurations are equal. The system thus approaches a stationary equilibrium state obeying detailed balance in which all interface configurations compatible with the restrictions $|h_i-h_{i+1}|=1$ and $h_i\geq 0$ occur with the same probability. Although in an infinite system this state is never reached, a well-defined quantity is the conditional correlation function
\begin{equation}
\label{corrfunction}
c(\ell) = \frac{\langle \delta_{h_i,0}\,\delta_{h_{i+\ell},0} \rangle}{\langle \delta_{h_i,0} \rangle}
\end{equation}
where $\langle\cdot\rangle$ denotes the spatial average over the index $i$. This correlation function may be interpreted as the conditional probability to find site $i+\ell$ at zero height given that site $i$ is a contact point, too. As the interface roughens, numerator and denominator tend to zero but their quotient tends to a well-defined non-zero value.

The present proof relies on the assumption that this two-point function -- like other two-point functions in scale-free situations -- decays according to the power law
\begin{equation}
c(\ell) \sim \ell^{-\theta z}\,,
\end{equation}
where $z=2$ is the dynamical exponent. As will be shown, the exponent $\theta$ can be computed by deriving an exact expression for $c(\ell)$.

Adopting a transfer matrix approach used in Ref.~\cite{Hilhorst,Burkhardt}, the correlation function~(\ref{corrfunction}) may be expressed as
\begin{equation}
\label{Cl}
C(\ell) = \frac{\langle 1 | T^\ell | 1\rangle}{\Lambda^\ell}\,,
\end{equation}
where $T$ is an infinite-dimensional transfer matrix with matrix elements
\begin{equation}
T_{h,h'}=\left\{
\begin{array}{ll}
1 & \mbox{ if } |h-h'|=1 \\
0 & \mbox{ otherwise }
\end{array}
\right.
\qquad \qquad
h,h' \in \{0,1,2,\ldots\}
\end{equation}
while $\Lambda=-2$ its dominating eigenvalue and $\langle 1|=|1\rangle^\dagger = (1,0,0,\ldots)$ denotes the vector representing height zero. Because of the special structure of $T$, where only the secondary diagonals are occupied, the evaluation of Eq.~(\ref{Cl}) reduces to counting all possible paths of the interface from site $i$ to site $i+\ell$. The number of such paths is given by Catalan numbers, leading to the exact solution
\begin{equation}
C(\ell) = 
\left\{
\begin{array}{ll}
\frac{\ell!}{(\ell/2)!\,(\ell/2+1)!\,2^\ell} & \mbox{ if } \ell \mbox{ even} \\
0 & \mbox{ if } \ell \mbox{ odd.}
\end{array}
\right.
\end{equation}
Inserting Stirling's formula $n!\simeq \sqrt{2\pi} n^{n+1/2} e^{-n}$ it straight forward to show that this correlation function decays asymptotically as
\begin{equation}
c(\ell) \simeq \sqrt{\frac{8}{\pi}}\,e\,\,\ell^{-3/2}
\end{equation}
Together with the known exponent $z=2$ we thus arrive at $\theta=3/4$.


\vspace{10mm}
\noindent
\large

\end{document}